\documentclass[
  aps,
  prb,
  reprint,
  superscriptaddress,
  amsmath,
  amssymb
]{revtex4-2}

\usepackage{graphicx}
\usepackage{bm}
\usepackage{siunitx}
\usepackage[hidelinks,hypertexnames=false]{hyperref}
\usepackage{dcolumn}
\usepackage{bm}
\usepackage[utf8]{inputenc}
\usepackage[T1]{fontenc}
\usepackage{lineno}
\usepackage{xcolor}
\usepackage{svg}
\usepackage{float}
\usepackage{algorithm}
\usepackage{algpseudocode}
\usepackage{dsfont}
\usepackage{physics}
\usepackage{flushend}
\usepackage{mathtools}
\usepackage{bbold}
\usepackage{color} 
\usepackage{upgreek}
\usepackage{tcolorbox}
\usepackage{soul}
\usepackage{booktabs}

\usepackage[normalem]{ulem} 

\newcommand{\add}[1]{}
\newcommand{\delete}[1]{}

\begin{document}

\preprint{APS/123-QED}

\title{Disorder signatures emerging at millikelvin temperatures \\in Si/SiGe field-effect stacks}

\author{Lino Visser}
\thanks{These authors contributed equally to this work.}
\affiliation{JARA-FIT Institute for Quantum Information, Forschungszentrum J\"ulich GmbH and RWTH Aachen University, Aachen, Germany}
\author{Alberto Mistroni}
\thanks{These authors contributed equally to this work.}
\affiliation{IHP - Leibniz Institute for High Performance Microelectronics, Frankfurt (Oder) 15236, Germany}
\author{Yuji Yamamoto}
\affiliation{IHP - Leibniz Institute for High Performance Microelectronics, Frankfurt (Oder) 15236, Germany}
\author{Fabian Fidorra}
\affiliation{IHP - Leibniz Institute for High Performance Microelectronics, Frankfurt (Oder) 15236, Germany}
\author{Oksana Fursenko}
\affiliation{IHP - Leibniz Institute for High Performance Microelectronics, Frankfurt (Oder) 15236, Germany}
\author{Steffen Marschmeyer}
\affiliation{IHP - Leibniz Institute for High Performance Microelectronics, Frankfurt (Oder) 15236, Germany}
\author{Marvin H. Zoellner}
\affiliation{IHP - Leibniz Institute for High Performance Microelectronics, Frankfurt (Oder) 15236, Germany}
\author{Giovanni Capellini}
\affiliation{IHP - Leibniz Institute for High Performance Microelectronics, Frankfurt (Oder) 15236, Germany}
\affiliation{Dipartimento di Scienze, Università Roma Tre, Roma 00146, Italy}
\author{Dominique Bougeard}
\affiliation{Fakultät für Physik, Universität Regensburg, Regensburg 93040, Germany}
\author{Vincent Mourik}
\affiliation{JARA-FIT Institute for Quantum Information, Forschungszentrum J\"ulich GmbH and RWTH Aachen University, Aachen, Germany}
\author{Marco Lisker}
\affiliation{IHP - Leibniz Institute for High Performance Microelectronics, Frankfurt (Oder) 15236, Germany}
\author{Felix Reichmann}
\email{reichmann@ihp-microelectronics.com}
\affiliation{IHP - Leibniz Institute for High Performance Microelectronics, Frankfurt (Oder) 15236, Germany}

\begin{abstract}
The performance and scalability of electron spin qubits based on gate-defined quantum-dots in undoped Si/SiGe field-effect stacks remain constrained by disorder originating from the gate stack. Its coupling to the quantum well can be reduced by increasing the Si quantum well depth, while electrostatic charge history, for example through interface-trap filling, can further modify the effective disorder landscape. Although such disorder is commonly benchmarked through mobility measurements using magnetotransport and Hall bar devices, dedicated investigations at millikelvin temperatures relevant for quantum-dot operation remain limited. Here, we use temperature-dependent magnetotransport on Hall bar shaped field-effect transistors to investigate how mobility-based disorder signatures depend on quantum-well depth and charge history from \(1.5~\mathrm{K}\) down to the millikelvin regime. We show that magnetotransport characterization at \(1.5~\mathrm{K}\) captures the dominant mobility improvement associated with reduced dielectric-interface coupling, but can underestimate disorder differences that emerge at millikelvin temperatures, particularly in the low-density regime. Our results therefore highlight that millikelvin magnetotransport characterization of Hall bar devices can provide additional insight for optimizing Si/SiGe field-effect stacks, particularly in the context of  gate-defined quantum dot spin qubits. 
\end{abstract}

\maketitle

\section{Introduction}

Undoped Si/SiGe field-effect stacks are a promising platform for scalable quantum processors based on spin qubits in gate-defined quantum dots \cite{SiSpinQubitsReview2025}, owing to long coherence times \cite{StruckCoherence2020,WangCoherence2024} and compatibility with semiconductor manufacturing \cite{NeyensIntel2024,George12Qubits2025}. In this technology, nanostructured gates electrostatically define quantum dots by locally modulating the two-dimensional carrier density in a buried tensile-strained Si quantum well (QW), enabling control of individual charge and spin states. Despite rapid progress \cite{Watson2Qubits2018,Philips6Qubits2022,George12Qubits2025}, scalability remains limited by material-related challenges, particularly disorder and valley splitting \cite{ScappucciReview2021,BurkardReview2023}. Disorder typically arises from both long-range electrostatic fluctuations associated with charges in the gate stack, dielectric interfaces, and surrounding barriers, and shorter-range disorder near the QW, such as interface roughness, alloy disorder, or residual impurities \cite{CulcerDisorder2009,HuangDisorder2023}. Suppressing these contributions is therefore essential for achieving more homogeneous electrostatic landscapes and narrower device-to-device variations.

A key practical challenge is to quantify disorder in a way that is both predictive of qubit operation and suitable for efficient stack optimization. Magnetotransport measurements on Hall bar field-effect transistors (HB-FETs) offer a complementary large-scale probe of the same two-dimensional electron gas (2DEG) used to define quantum dots \cite{WuetzMagneto2020,EspostiMagneto2024}. Temperature-dependent magnetotransport can provide additional insight because cooling changes screening and charge dynamics, allowing transport features that appear similar at a fixed, elevated temperature to become distinguishable in the qubit-relevant millikelvin regime \cite{MiScattering2015}. This distinction is particularly relevant across carrier-density regimes: low-density transport is strongly influenced by long-range potential fluctuations, often associated with charges near the semiconductor--dielectric interface, whereas higher-density mobility becomes increasingly sensitive to short-range disorder close to the QW, for example interface roughness and background impurities \cite{HuangQWDepth2014,HuangDisorder2023}.

The depth of the strained-Si QW therefore represents a key design parameter for controlling the disorder landscape. Increasing the QW depth is expected to reduce coupling to gate-stack-related disorder \cite{HuangQWDepth2012}, but also weakens the gate tunability of the potential in the QW. This trade-off is directly relevant for quantum-dot gate-stack design, where disorder suppression must be balanced against sufficient electrostatic control. Furthermore, changing the QW depth may also shift the relative importance of long- and short-range disorder contributions.  Previous studies have investigated related trends, but were generally limited to heterostructures with lower Ge concentrations, thicker QWs, a narrower QW-depth range, and/or a more limited characterization-temperature window \cite{LuQWDepth2011,HuangQWDepth2012,LuQWDepth2013,HuangQWDepth2014,LarocheQWDepth2015,MiScattering2015,Su2DEG2019,WuetzMagneto2020}.

The charge environment itself can also depend on the electrostatic history of the device. Gate-voltage hysteresis is widely observed in semiconductor heterostructures and is commonly attributed to charge trapping and detrapping at or near the dielectric/semiconductor interface. While such hysteresis is often considered a complication for device tuning, recent work has shown that the charge state of the gate stack can also be controlled deliberately. Examples include optical illumination at cryogenic temperature, which can reproducibly shift threshold voltages in Si/SiGe quantum devices \cite{WolfeIllumination2024}, and biased cooldown protocols \cite{FerreroBiasCooling2024,DiebelBiasCooling2025}, which can modify the electrostatic operating point and reduce charge noise in gated Si/SiGe devices. In addition, Meyer et al. demonstrated that stress–voltage conditioning can permanently tune quantum-dot operating voltages, yielding stable shifts of hundreds of millivolts and reducing device-to-device variability in quantum-dot arrays \cite{MeyerUniformityDots2023}. Together, these studies show that interface charge states are not merely uncontrolled disorder sources, but can constitute an experimentally tunable part of the electrostatic landscape. For magnetotransport-based stack characterization, this raises the broader question of whether an intentionally charged state only changes threshold voltages and local turn-on conditions, or whether it also modifies the density-dependent mobility and its temperature dependence.

In order to quantify how low-temperature mobility in undoped Si/SiGe 2DEGs is shaped by both the device state and the heterostructure geometry, we combine in this work temperature-dependent magnetotransport, electrical charge-history control, and systematic variation of the electrostatic stack design. Using HB-FETs, we extract density-dependent mobility and percolation density from \(1.5~\mathrm{K}\)  down to millikelvin temperatures, and use the fractional mobility change upon cooling to compare the temperature-sensitive part of the transport across density regimes. We first identify the temperature-sensitive transport regimes in a \(37~\mathrm{nm}\) reference stack, then examine how charging modifies mobility and device homogeneity, and finally compare QWs reaching a depth of 160 nm to determine how electrostatic stack design changes the low-temperature transport response. We show that mobility-density characteristics can be substantially more sensitive at millikelvin temperatures to the stack design and device charge state, particularly in the low-density regime.

\section{Results}

\subsection{Identifying temperature-sensitive transport regimes}

We first examine how the transport characteristics of the reference Si/SiGe field-effect stack evolve upon cooling into the millikelvin regime. For this purpose, we analyze a reference device comprising a \(10~\mathrm{nm}\) thick Si oxide as the gate dielectric and a \(37~\mathrm{nm}\) deep, \(8~\mathrm{nm}\) thick Si QW \cite{Mistroni2DEG2025}. To minimize the influence of persistent charge configurations, the device was warmed up to room temperature for \(30~\mathrm{min}\) after each temperature point before being cooled down again for the subsequent measurement, showing consistent turn-on behavior after each reset. This pristine device state then allows us to compare the temperature evolution of the density-dependent mobility and percolation threshold without interference from charge-history effects.

\begin{figure}
    \centering
    \includegraphics[width=1\linewidth]{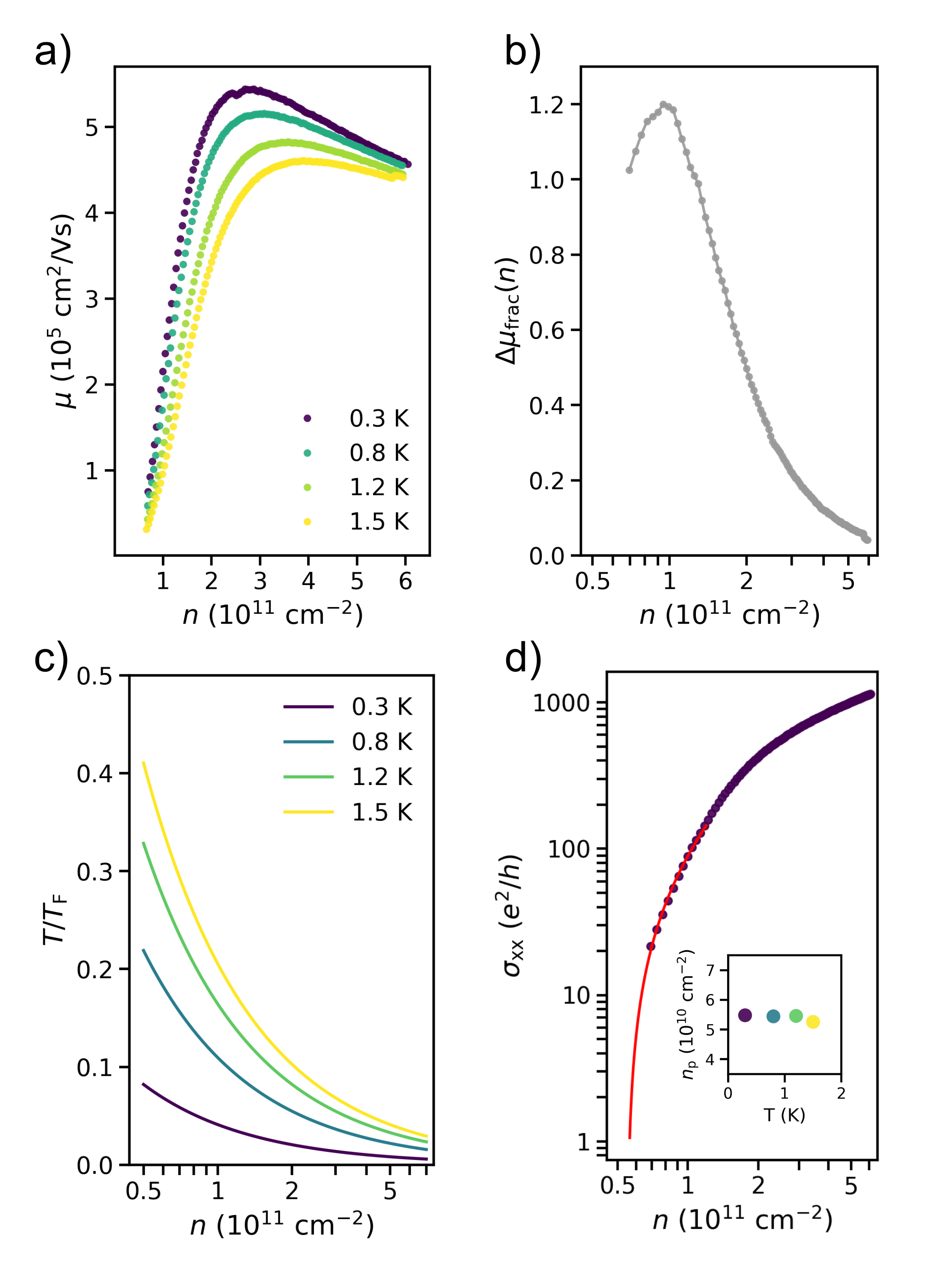}
    \caption{Temperature-dependent transport of the pristine reference Si/SiGe field-effect stack with a \(37~\mathrm{nm}\) deep and 10 nm thick QW. a) Hall mobility \(\mu\) as a function of carrier density \(n\), measured at \(1.5~\mathrm{K}\), \(1.2~\mathrm{K}\), \(0.8~\mathrm{K}\) and \(0.3~\mathrm{K}\). b) Fractional mobility gain \(\Delta\mu_{\mathrm{frac}}(n)\) upon cooling from \(1.5~\mathrm{K}\) to \(0.3~\mathrm{K}\). c) Calculated ratio \(T/T_F\) as a function of density for the measured temperatures, using the same color code as in b). d) Longitudinal conductivity \(\sigma_{xx}\) as a function of density at \(0.3~\mathrm{K}\). The red line is the percolation fit, used to extract the percolation threshold \(n_p\). The inset shows extracted \(n_p\) values at \(1.5~\mathrm{K}\), \(1.2~\mathrm{K}\), \(0.8~\mathrm{K}\) and \(0.3~\mathrm{K}\) (same color code as a) and c)).
}
    \label{Figure1}
\end{figure}

Figure~\ref{Figure1}(a) shows the Hall mobility \(\mu\) as a function of carrier density \(n\) at \(1.5~\mathrm{K}\), \(1.2~\mathrm{K}\), \(0.8~\mathrm{K}\), and \(0.3~\mathrm{K}\). As the temperature decreases, the mobility increases over a broad density range and the mobility maximum shifts from approximately \(3.5\times10^{11}~\mathrm{cm^{-2}}\) at \(1.5~\mathrm{K}\) to about \(2.5\times10^{11}~\mathrm{cm^{-2}}\) at \(0.3~\mathrm{K}\).  Cooling into the sub-Kelvin regime therefore not only rescales the absolute mobility, but also changes its density-dependent shape. This becomes particularly evident when comparing the low- and high-density regimes. In the carrier density regime above \(5\times10^{11}~\mathrm{cm^{-2}}\), the mobility barely changes when cooling  below \(1.5~\mathrm{K}\), while the density regime below approximately \(3\times10^{11}~\mathrm{cm^{-2}}\) remains temperature sensitive down to \(0.3~\mathrm{K}\). 

To  quantify the density-dependent low-temperature mobility response, Fig.~\ref{Figure1}(b) shows the fractional mobility gain upon cooling from \(1.5~\mathrm{K}\) to \(0.3~\mathrm{K}\),
\[
\Delta\mu_{\mathrm{frac}}(n)
=
\frac{\mu(0.3~\mathrm{K},n)-\mu(1.5~\mathrm{K},n)}
{\mu(1.5~\mathrm{K},n)}.
\]
The fractional gain first increases with carrier density, reaches a maximum at  \(1\times10^{11}~\mathrm{cm^{-2}}\), and then decreases for higher density, confirming that the sub-\(1.5~\mathrm{K}\) mobility enhancement is not uniform across density.

To interpret this density-dependent temperature response, we compare the measured transport regime with the relevant degeneracy scale of the 2DEG. Within the effective-mass approximation for a parabolic 2D electron system, the Fermi temperature is defined as \(T_F=E_F/k_B\), where \(E_F\) is the Fermi energy and \(k_B\) is the Boltzmann constant. The Fermi wave vector is given by \(k_F=(4\pi n/g_sg_v)^{1/2}\), where \(n\) is the carrier density and \(g_s\) and \(g_v\) are the spin and valley degeneracies, respectively \cite{DasSarma2DEG2015}. For this qualitative estimate, we treat the two low-lying valleys of the Si/SiGe 2DEG as effectively degenerate and use \(g_s=2\) and \(g_v=2\), together with the in-plane effective mass \(m^*\approx0.19\,m_0\). This yields
\[
T_F=\frac{\pi \hbar^2 n}{2 k_B m^*}.
\]
The ratio \(T/T_F\) therefore provides a dimensionless estimate of the degeneracy of the electron system, with \(T/T_F \ll 1\) corresponding to a strongly degenerate 2DEG. Based on these assumptions, Figure~\ref{Figure1}(c) shows the calculated \(T/T_F\) as a function of carrier density for each temperature used in the mobility measurements reported in Fig.~\ref{Figure1}(a). 

Across all measurement temperatures, \(T/T_F\)  decreases with increasing carrier density. Consequently, the high-density regime is characterized by a uniformly small \(T/T_F\) ratio , indicating that the 2DEG remains strongly degenerate throughout the investigated temperature range. Within screening-based descriptions of low-temperature 2D transport, the temperature dependence arising from screened Coulomb disorder becomes weak in this low-\(T/T_F\) regime \cite{DasSarma2DEG2015}. This is consistent with the weak residual temperature dependence of the high-density mobility observed in Fig.~\ref{Figure1}(a) and (b). In addition, temperature-insensitive mechanisms such as interface roughness and alloy disorder are expected to play an increasingly important role at high carrier density, further reducing the relative impact of temperature-dependent screening. Towards lower carrier density, \(T_F\) decreases and the absolute separation in \(T/T_F\)between the different measurement temperatures becomes increasingly pronounced. Consequently, changing the measurement temperature corresponds to a larger change relative to the degeneracy scale of the 2DEG. Temperature-dependent screening of long-range Coulomb disorder can therefore have a stronger influence on transport in this regime \cite{DasSarma2DEG2015}, consistent with the more pronounced mobility enhancement observed upon cooling at lower carrier density. 

However, the non-monotonic density dependence of \(\Delta\mu_{\mathrm{frac}}(n)\) shows that the mobility gain is not governed by \(T/T_F\) alone. In a simple degeneracy-based picture, the temperature sensitivity would be expected to decrease monotonically with increasing density because \(T_F\propto n\). Instead, the strongest response occurs at a carrier density of around \(1\times10^{11}~\mathrm{cm^{-2}}\), where the system is already conducting but the screening of long-range potential fluctuations remains comparatively weak and temperature-dependent. This behavior is consistent with screening-based descriptions of low-density two-dimensional transport, in which temperature-dependent screening of long-range Coulomb disorder can produce the strongest temperature dependence at intermediate densities above the critical density \cite{DasSarmaLowDensity1999}. In this picture, the response is suppressed near turn-on by disorder broadening and inhomogeneous or percolative transport \cite{DasSarmaLowDensity2005}, and reduced again at high density as \(T_F\) increases \cite{DasSarmaLowDensity1999,DasSarma2DEG2015}.

As a final consistency check, we examine whether the low-density conduction onset shifts over the same temperature range. Figure~\ref{Figure1}(d) shows a representative fit of the low-density longitudinal conductivity \(\sigma_{xx}\) at \(0.3~\mathrm{K}\) to the percolation form
\[
\sigma_{xx} \propto (n-n_p)^p,
\]
where \(n_p\) is used as the critical density for metallic conduction. The inset shows that the extracted \(n_p\) remains essentially unchanged between \(1.5~\mathrm{K}\) and \(0.3~\mathrm{K}\). Since \(n_p\) reflects the low-density conduction onset and is sensitive to long-range potential fluctuations and density inhomogeneity, a substantial change in the effective electrostatic disorder landscape would be expected to shift the extracted threshold. The absence of a measurable shift in \(n_p\) therefore suggests that the observed mobility enhancement is not caused by a change in the percolative threshold, but instead arises from changes in scattering and screening. Consistently, the Hall density \(n(V_{\mathrm{G}})\) remains linear in gate voltage and shows no detectable temperature-dependent change in slope or voltage offset over the same temperature range (see Supplemental Material).

Taken together, these results identify the \(1.5~\mathrm{K}\) to \(0.3~\mathrm{K}\) comparison as a sensitive probe of low-density disorder and screening in the millikelvin regime. The pristine reference measurement establishes the baseline for the following section, where we examine how intentional electrostatic conditioning modifies the temperature-dependent mobility response.

\begin{figure*}[t]
    \centering
    \includegraphics[width=0.8\linewidth]{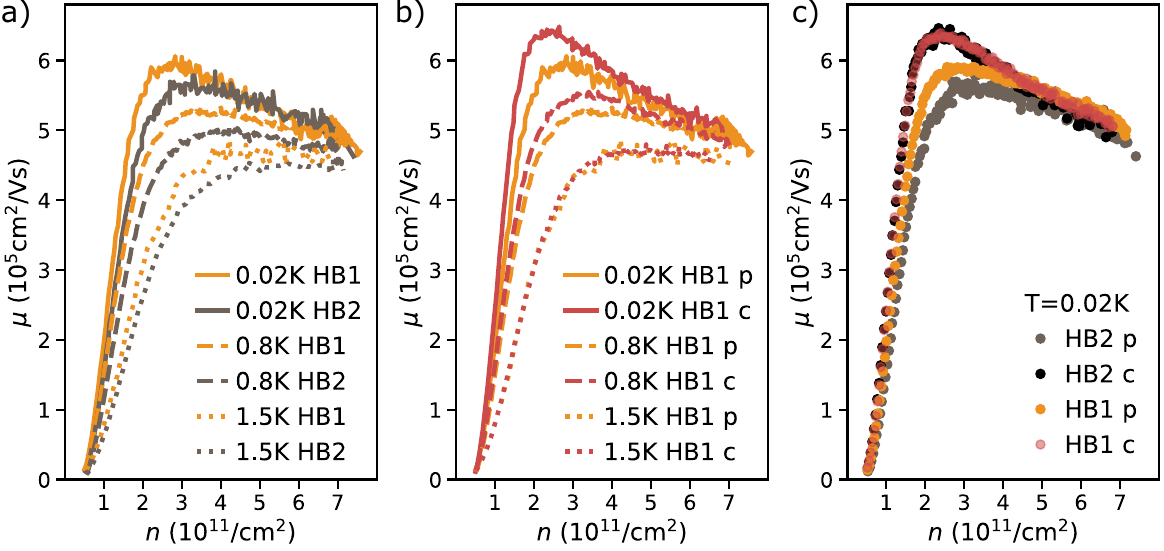}
    \caption{Charge-history-dependent mobility in two HB-FETs (HB1 and HB2) from the \(37~\mathrm{nm}\)-deep QW reference stack. a) Mobility as a function of carrier density for HB1 (orange) and HB2 (grey) in the pristine state, measured at \(1.5~\mathrm{K}\) (dotted lines), \(0.8~\mathrm{K}\) (dashed lines), and \(0.02~\mathrm{K}\) (solid lines). b) Comparison of pristine and charged-state mobility curves for HB1. Pristine curves are labeled ``p'' and shown in orange, while charged-state curves are labeled ``c'' and shown in red. The line styles denote the measurement temperature as in panel a). c) Comparison of HB1 and HB2 at \(0.02~\mathrm{K}\) before and after charging. Pristine states are labeled HB1 p and HB2 p and shown in orange and grey, respectively, while charged states are labeled HB1 c and HB2 c and shown in red and black, respectively.}
    \label{Figure2}
\end{figure*}

\subsection{Charge-history-dependent mobility at millikelvin temperatures}

We next investigate how bias charge history affects the low-temperature mobility of devices from the 37 nm reference stack. For this purpose, two HB-FETs, denoted as HB1 and HB2, were taken from the same wafer as the reference device discussed in the previous section \cite{Mistroni2DEG2025}. In addition to measurements at 1.5 K and 0.8 K, the devices were characterized down to 0.02 K in order to probe the transport response in a temperature regime closer to typical spin-qubit operation. The measurements distinguish between a pristine (labeled as "p") and a charged state (labeled as "c"). The pristine state corresponds to the first measurement after cool down, before the device has been exposed to large positive top-gate voltages. During this measurement, the gate voltage was stepped up to \(V_\mathrm{G}=1.5~\mathrm{V}\), while the magnetic field was swept at each gate-voltage setpoint to extract the carrier density and mobility. At high positive gate bias, the carrier density no longer follows the linear capacitive \(n(V_\mathrm{TG})\) response, but instead enters a saturation regime (see Supplemental Material.). This behavior indicates that the applied gate voltage exceeds the purely capacitive operating range and induces charge tunneling or charge rearrangement in the gate stack \cite{LuQWDepth2011}. Subsequent measurements at the same temperature are therefore referred to as charged device measurements. 

Figure~\ref{Figure2}(a) first compares the pristine mobility-density characteristics of HB1 (orange) and HB2 (grey) at 1.5 K (dotted lines), 0.8 K (dashed lines), and 0.02 K (solid lines). Both devices show the same qualitative temperature dependence as the reference stack in the previous section. This confirms that the two devices are representative of the same material stack and reproduce the temperature-dependent transport response. At the same time, HB1 exhibits a systematically higher mobility than HB2 in the pristine state, attributed to device-to-device variability, as we reported previously \cite{Mistroni2DEG2025}.

\begin{figure*}[t]
    \centering
    \includegraphics[width=0.85\textwidth]{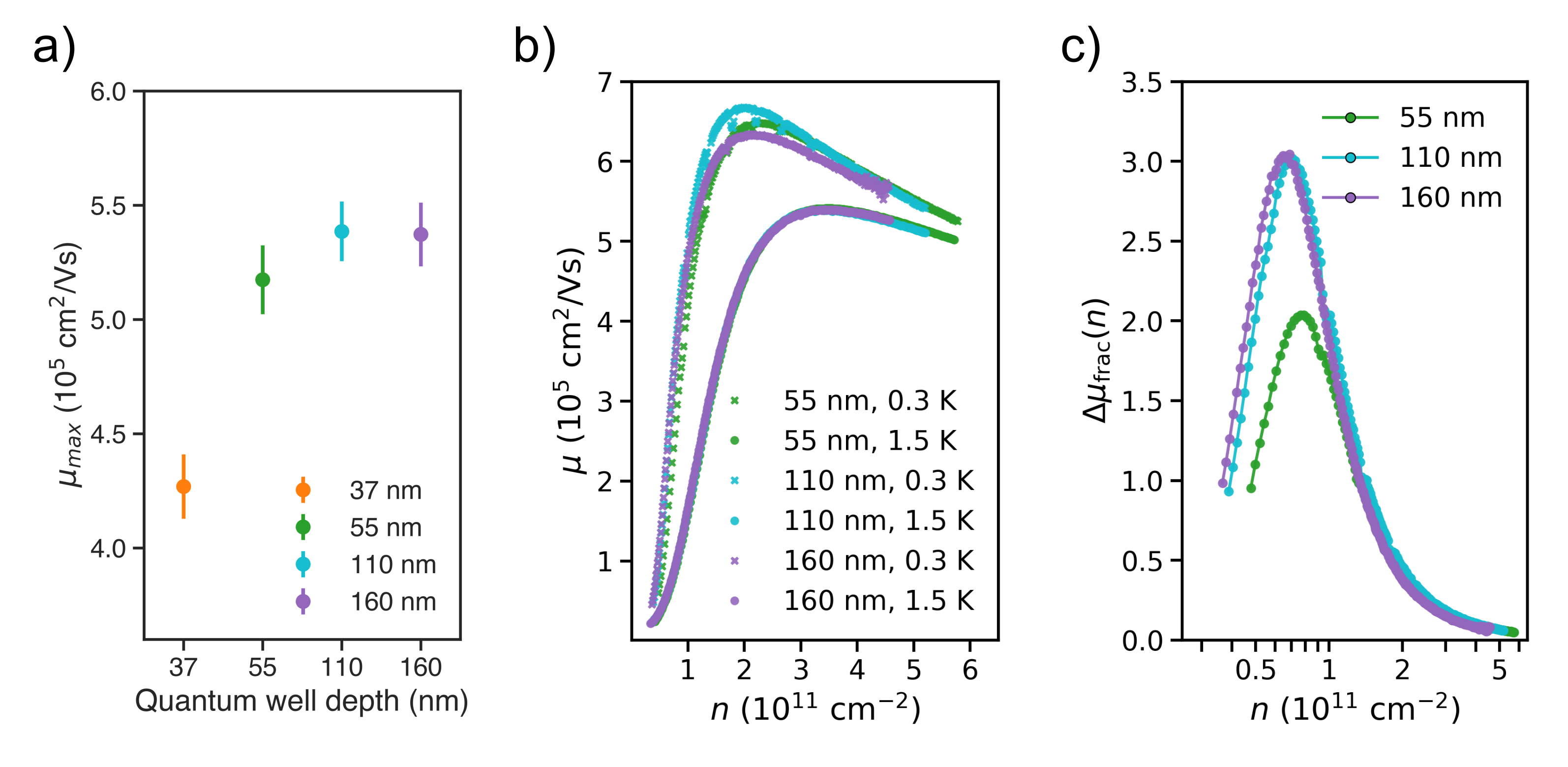}
    \caption{
 QW-depth dependence of the \(1.5~\mathrm{K}\) mobility benchmark and millikelvin mobility response.
    a) Maximum mobility \(\mu_{\max}\), extracted from \(1.5~\mathrm{K}\) mobility traces, as a function of QW depth for \(37~\mathrm{nm}\) (orange, 10 samples measured), \(55~\mathrm{nm}\) (green, 5 samples measured), \(110~\mathrm{nm}\) (cyan, 5 samples measured), and \(160~\mathrm{nm}\) (purple, 5 samples measured) stacks. Each dot represents the mean, while the error bar indicates the standard deviation. 
    b) Hall mobility \(\mu\) as a function of carrier density \(n\) for representative devices from the \(55~\mathrm{nm}\) (green), \(110~\mathrm{nm}\) (cyan), and \(160~\mathrm{nm}\) (purple) stacks, measured at \(1.5~\mathrm{K}\) (dots) and \(0.3~\mathrm{K}\) (crosses).
    c) Fractional mobility gain \(\Delta\mu_{\mathrm{frac}}(n)\) upon cooling from \(1.5~\mathrm{K}\) to \(0.3~\mathrm{K}\) for the representative devices shown in b) with the same color code.
    }
    \label{Figure3}
\end{figure*}

To isolate the influence of charge history, Fig.~\ref{Figure2}(b) compares mobility curves of HB1 in pristine (labeled as "p", orange curves) and charged device states (labeled as "c", red curves) at different temperatures, indicated by the same line style as in Figure~\ref{Figure2}(a). At \(1.5~\mathrm{K}\), the mobility is essentially unchanged after the high-bias sweep leading to a charged device, indicating that conventional \(1.5~\mathrm{K}\) transport screening is largely insensitive to this charge-history effect. At lower temperatures, however, the charged device state exhibits a clear mobility enhancement. The effect is strongest at \(0.02~\mathrm{K}\) and is most pronounced in the intermediate-density regime around \(2\times10^{11}~\mathrm{cm^{-2}}\), suggesting that the high-bias sweep does not only shift the threshold voltage, but modifies a disorder contribution that becomes relevant only in the sub-Kelvin regime.

This behavior is consistent with a partial filling or rearrangement of charge traps near the dielectric/semiconductor interface \cite{Su2DEG2019}. In this picture, electrons tunneling at large positive gate bias modify the local electrostatic conditions of the channel. Such a change is expected to be most visible in the low- to intermediate-density regime, where the mobility is particularly sensitive to long-range Coulomb scattering from remote charged centers. The resulting charge configuration can therefore reduce the strength of these long-range potential fluctuations, thereby increasing the mobility in this density range. The absence of a significant mobility change below \(1.5~\mathrm{K}\) at higher carrier density, here above approximately \(4\times10^{11}~\mathrm{cm}^{-2}\), further supports this interpretation. In this regime, transport is expected to be less sensitive to remote-charge disorder and more strongly limited by scattering mechanisms such as interface roughness or alloy disorder.

Finally, Fig.~\ref{Figure2}(c) compares the pristine and charged device states of HB1 and HB2 at \(0.02~\mathrm{K}\), where the charge-history effect is most apparent. In the pristine state, the 20 mK mobility curves of HB1 and HB2, replotted from Fig.~\ref{Figure2}(a), show a clear device-to-device difference, with the largest separation occurring around \(n \approx 2\times10^{11}~\mathrm{cm}^{-2}\). After charging, however, the mobility curves of HB1 and HB2 overlap across the full measured density range. This convergence indicates that the high-bias charge history suppresses a device-specific component of the low-temperature disorder landscape. In other words, part of the pristine mobility variation between nominally identical devices appears to originate from charge configurations that can be modified by high-gate-voltage operation.

These results show that charge history can remain largely hidden in \(1.5~\mathrm{K}\) characterization, while becoming visible in millikelvin transport. The charged device state not only enhances the low-temperature mobility, but also reduces the device-to-device spread between nominally identical HB-FETs. This highlights the importance of controlling and documenting electrostatic history when using Hall bar transport as a diagnostic for qubit-relevant heterostructures and gate stacks in the millikelvin regime. In the following section, we therefore extend this temperature-dependent analysis to pristine devices with deeper QWs, to estimate intrinsic stack-design trends.

\subsection{QW-depth dependence of the millikelvin mobility response}

We next investigate how increasing the QW depth affects the mobility response from conventional \(1.5~\mathrm{K}\) screening down to the millikelvin regime. As a starting point, we compare the maximum mobility at \(1.5~\mathrm{K}\) across the full QW-depth design series, with top SiGe barrier thicknesses of \(37\), \(55\), \(110\), and \(160~\mathrm{nm}\), while keeping the barrier Ge concentration and QW thickness fixed at \(33\%\) and \(8~\mathrm{nm}\), respectively. As described in the methods section, the \(55\), \(110\), and \(160~\mathrm{nm}\) stacks were fabricated with a nominal \(30~\mathrm{nm}\) gate dielectric, whereas the \(37~\mathrm{nm}\) reference stack used a \(10~\mathrm{nm}\) gate dielectric. We therefore use this full-series comparison only as a compact \(1.5~\mathrm{K}\) benchmark of the overall depth-dependent design trend. To compare the millikelvin mobility response under matched nominal gate-stack conditions, we then restrict the paired \(1.5~\mathrm{K}\)--\(0.3~\mathrm{K}\) analysis to representative devices from the \(55\), \(110\), and \(160~\mathrm{nm}\) stacks.

Figure~\ref{Figure3}(a) summarizes the \(1.5~\mathrm{K}\) maximum mobility \(\mu_{\max}\) across the full QW-depth design series, comprising \(37~\mathrm{nm}\) (orange, \(10\) samples), \(55~\mathrm{nm}\) (green, \(5\) samples), \(110~\mathrm{nm}\) (cyan, \(5\) samples), and \(160~\mathrm{nm}\) (purple, \(5\) samples) stacks. The largest change is observed between the \(37~\mathrm{nm}\) reference stack and the \(55~\mathrm{nm}\) stack, where \(\mu_{\max}\) increases by more than \(20\%\). This trend is consistent with reduced electrostatic coupling of the 2DEG to disorder sources near the dielectric interface as the QW is moved farther from the surface \cite{HuangQWDepth2012}. Further increasing the QW depth to \(110~\mathrm{nm}\) yields only a moderate additional gain, while the \(160~\mathrm{nm}\) stack does not show a further systematic improvement within the device-to-device spread. Thus, within the present material and gate-stack platform, the \(1.5~\mathrm{K}\) benchmark indicates that the dominant mobility improvement occurs when moving from the shallow \(37~\mathrm{nm}\) reference design to the \(55~\mathrm{nm}\) stack, whereas substantially deeper wells provide diminishing returns in \(\mu_{\max}\). 

However, \(\mu_{\max}\) is a single benchmark value and does not capture how the full \(\mu(n)\) response evolves upon cooling toward the qubit-relevant regime. As shown above for the \(37~\mathrm{nm}\) reference stack, the mobility can continue to change below \(1.5~\mathrm{K}\), particularly at low carrier density. We therefore compare paired \(1.5~\mathrm{K}\) and \(0.3~\mathrm{K}\) mobility traces of additional, randomly selected devices with  \(55\), \(110\), and \(160~\mathrm{nm}\) deep QWs, as shown in Fig.~\ref{Figure3}(b).  At \(1.5~\mathrm{K}\), the mobility traces overlap over a broad density range. At  \(0.3~\mathrm{K}\) however, the differences in mobility between the devices become more pronounced. The clearest separation appears in the intermediate-density range from approximately \(1\times10^{11}\) to \(3\times10^{11}~\mathrm{cm^{-2}}\), i.e. around the mobility maximum. In this regime, the \(110~\mathrm{nm}\) stack exhibits the highest mobility out of all devices, while the \(160~\mathrm{nm}\) stack shows a lower maximum mobility  than the \(55~\mathrm{nm}\) device. This example highlights that in the intermediate-density regime, the maximum mobility  becomes increasingly sensitive to the balance between residual long-range disorder and sample-specific internal disorder contributions, such as background charged impurities and short-range interface-related scattering, that persist after the dominant interface-related scattering has been suppressed. At carrier densities above \(3\times10^{11}~\mathrm{cm^{-2}}\), the mobility curves begin to converge again. This is consistent with the interpretation developed for the reference sample: in the high-density regime, transport is increasingly governed by short-range and largely temperature-independent scattering mechanisms such as interface roughness and possibly alloy disorder. Since all three heterostructures share nominally the same Ge concentration, QW thickness, and Si/SiGe interfaces, these high-density scattering contributions are not expected to differ strongly between the devices.

To quantify where the mobility changes most strongly upon cooling, we evaluate the fractional mobility gain \(\Delta\mu_{\mathrm{frac}}(n)\) introduced in the first section. As shown in Fig.~\ref{Figure3}(c), all heterostructures follow the same qualitative trend already observed in the \(37~\mathrm{nm}\) reference stack, with \(\Delta\mu_{\mathrm{frac}}(n)\) increasing with carrier density, reaching a pronounced maximum around \(0.7\times10^{11}~\mathrm{cm}^{-2}\), and then decreasing again toward higher density. In this low-density regime, the fractional gain follows the expected QW-depth trend, with the smallest response for the \(55~\mathrm{nm}\) stack and a larger, nearly saturated enhancement for the \(110~\mathrm{nm}\) and \(160~\mathrm{nm}\) stacks. This suggests that the temperature-sensitive part of the mobility response remains broadly consistent with reduced coupling to remote disorder in deeper QWs. By contrast, the absolute \(\mu(n)\) traces at \(0.3~\mathrm{K}\), shown in Fig.~\ref{Figure3}(b), and in particular the mobility maxima, do not follow a simple monotonic ordering with QW depth, since the \(110~\mathrm{nm}\) device exhibits the highest maximum mobility, whereas the \(160~\mathrm{nm}\) device remains below the \(55~\mathrm{nm}\) device. This indicates that the millikelvin mobility maximum is not governed by QW--dielectric separation alone, but is additionally sensitive to sample-specific disorder contributions, potentially including heterostructure-internal charged impurities, local interface roughness variations, or alloy-related scattering.

This section has shown that in the \(1.5~\mathrm{K}\) benchmark the dominant increase in \(\mu_{\max}\) occurs between the \(37~\mathrm{nm}\) and \(55~\mathrm{nm}\) stacks, with diminishing returns for deeper QWs. Paired \(1.5~\mathrm{K}\)--\(0.3~\mathrm{K}\) measurements revealed a pronounced density-dependent mobility enhancement upon cooling. This response is strongest in the low-to-intermediate density regime and is more clearly captured by the fractional mobility gain than by \(\mu_{\max}\) alone. The non-monotonic ordering of the absolute mobility maxima at \(0.3~\mathrm{K}\) further suggests that millikelvin mobility is additionally sensitive to heterostructure-internal or sample-specific variations once the impact of the dielectric interface is reduced.

\section{Conclusion}

In conclusion, we investigated the interplay of QW depth, charge history, and characterization temperature in undoped Si/SiGe field-effect stacks using magnetotransport on HB-FETs. Benchmarking at \(1.5~\mathrm{K}\) shows a strong initial improvement in mobility when the QW depth is increased from \(37~\mathrm{nm}\) to \(55~\mathrm{nm}\), followed by only small additional gains for deeper wells. This indicates that, in the present materials platform, moderate QW depths already strongly suppress the dominant dielectric-interface-related disorder contribution, providing a useful first-order design guideline for future quantum-dot devices.

Temperature-dependent measurements show, however, that this first-order optimization does not fully capture the residual low-density disorder response that emerges under millikelvin measurement conditions, which are more representative of the effective operating regime of quantum-dot devices. While the high-density mobility approaches its low-temperature limit near \(1.5~\mathrm{K}\), the low-density regime remains strongly temperature-dependent down to \(0.3~\mathrm{K}\). In particular, the cooling-induced mobility enhancement is not a uniform rescaling of \(\mu(n)\), but is strongest in the low-to-intermediate density regime as revealed by \(\Delta\mu_{\mathrm{frac}}(n)\). The observed non-monotonic density dependence cannot be explained by a \(T/T_F\)-based picture alone, for which the temperature sensitivity would be expected to decrease continuously with increasing density. Instead, it is consistent with a screening-based description of low-density 2D transport, in which the temperature dependence is enhanced near the onset of metallic conduction and then weakens as the carrier density increases and screening becomes more effective.

The charged-device measurements presented in this work provide a complementary test of the observed low-density temperature sensitivity. At \(1.5~\mathrm{K}\), the pristine and charged \(\mu(n)\) traces are nearly identical, whereas clear differences emerge upon cooling below \(1.5~\mathrm{K}\). These differences are concentrated in the low-to-intermediate density regime, while the high-density mobility converges across charging states at all temperatures, consistent with the reduced temperature sensitivity expected once carrier density and screening increase. Moreover, at the lowest measurement temperature of \(0.02~\mathrm{K}\), devices that show pronounced device-to-device variations in the pristine state exhibit overlapping \(\mu(n)\) traces after charging. This suggests that controlled charge filling can provide a practical route to smooth the effective electrostatic potential landscape.

Having established that the low-density millikelvin response is sensitive both to temperature and charge configuration, the deep-QW series shows how this response evolves with heterostructure design. The fractional mobility gain follows the expected QW-depth trend, with the smallest response for the \(55~\mathrm{nm}\) stack and a larger, nearly saturated enhancement for the \(110~\mathrm{nm}\) and \(160~\mathrm{nm}\) stacks. This suggests that the temperature-sensitive part of the mobility response remains broadly consistent with reduced coupling to remote disorder in deeper QWs. By contrast, the absolute millikelvin \(\mu(n)\) traces, and in particular the mobility maxima at \(0.3~\mathrm{K}\), do not follow a simple monotonic ordering with QW depth. This indicates that the millikelvin mobility maximum is not governed by QW--dielectric separation alone, but is additionally sensitive to sample-specific disorder contributions, potentially including heterostructure-internal charged impurities, local interface-roughness variations, or alloy-related scattering. The fractional mobility gain and the full millikelvin \(\mu(n)\) traces therefore provide complementary information. The former captures the systematic temperature-sensitive response to QW-depth-dependent remote disorder, while the latter can reveal residual disorder contributions that are obscured in conventional \(1.5~\mathrm{K}\) mobility benchmarks.

Overall, these results show that conventional \(1.5~\mathrm{K}\) mobility benchmarking provides an important but incomplete basis for Si/SiGe stack optimization. Once the dominant dielectric-interface contribution is reduced, the residual disorder landscape becomes most clearly visible through density-dependent millikelvin transport, charge-history sensitivity, and the comparison between fractional mobility gain and absolute \(\mu(n)\) traces. Millikelvin HB-FET characterization therefore provides a complementary and experimentally accessible benchmark for identifying disorder limitations that are relevant for the continued optimization of Si/SiGe field-effect stacks for quantum-dot and spin-qubit devices.

\section{Methods}

\subsection{Heterostructure Growth}

All Si/SiGe heterostructures reported in this work were grown by industry-standard reduced-pressure chemical vapor deposition \cite{Mistroni2DEG2025}. The starting substrates were \(200~\mathrm{mm}\), p-doped Si(100) wafers with a resistivity of \(5\text{--}22~\Omega\mathrm{cm}\). After a pre-epitaxy wafer clean, a \(4~\mu\mathrm{m}\)-thick step-graded Si$_{1-x}$Ge$_x$ virtual substrate was grown, terminating at a nominal Ge concentration of \(x = 0.33 \pm 0.01\). On top of this virtual substrate, a \(2.7~\mu\mathrm{m}\)-thick constant-composition Si$_{0.67}$Ge$_{0.33}$ buffer layer was deposited. Chemical-mechanical polishing was then applied to reduce the surface roughness associated with cross-hatch formation. After a second surface clean, a \(100~\mathrm{nm}\)-thick Si$_{0.67}$Ge$_{0.33}$ bottom barrier was grown, followed by an \(8~\mathrm{nm}\)-thick strained-Si quantum well and a Si$_{0.67}$Ge$_{0.33}$ top barrier with nominal thickness of \(37\), \(55\), \(110\), or \(160~\mathrm{nm}\). These layers were grown at \(600\,^\circ\mathrm{C}\). Finally, the heterostructure was terminated by a \(6~\mathrm{nm}\)-thick epitaxial sacrificial Si cap grown at \(700\,^\circ\mathrm{C}\), which protects the underlying SiGe barrier from native oxidation.

\subsection{Device Fabrication}

Following heterostructure growth, Hall bar field-effect transistors (HB-FETs) were fabricated using CMOS-compatible processing \cite{ReichmannHBFET2024}. Ohmic contacts were formed by selective phosphorus ion implantation. Phosphorus implantation was performed with a dose of \(4.5\times10^{15}~\mathrm{cm}^{-2}\) per implantation step. All QW depths received a \(20~\mathrm{keV}\) implant. Additional implantation energies were used for deeper QWs: \(40~\mathrm{keV}\) for the \(55~\mathrm{nm}\) QW, \(60~\mathrm{keV}\) for the \(110~\mathrm{nm}\) QW, and \(60\) and \(90~\mathrm{keV}\) for the \(160~\mathrm{nm}\) QW. After implantation, the wafers were annealed for \(1~\mathrm{min}\) at \(700\,^\circ\mathrm{C}\) to activate the phosphorus dopants. A high-density-plasma (HDP) SiO$_2$ layer was then deposited at \(300\,^\circ\mathrm{C}\) and served as the gate dielectric with a thickness of \(10~\mathrm{nm}\) on top of the 37 nm deep QWs and \(30~\mathrm{nm}\) on top of the 55, 110 and 160 nm deep QWs. During fabrication, the sacrificial Si cap was gradually consumed by repeated native-oxide formation and subsequent HF dips, and was fully removed during HDP SiO$_2$ deposition, as confirmed by in-line ellipsometry. A TiN top-gate layer was subsequently deposited by physical vapor deposition (PVD), patterned into the Hall bar geometry by optical lithography, and etched by reactive ion etching. The fabricated HB-FETs have a channel length of \(2~\mathrm{mm}\), a channel width of \(20~\mu\mathrm{m}\), and a voltage-probe spacing of \(300~\mu\mathrm{m}\).

\subsection{Magnetotransport Characterization}

Temperature-dependent measurements of the reference stack and the deeper-QW devices were performed in a Kiutra L-Type rapid cryostat over the temperature range from \(4~\mathrm{K}\) down to \(0.3~\mathrm{K}\). The \(1.5~\mathrm{K}\) benchmarking measurements of the QW-depth series were performed separately in an Oxford Instruments Teslatron-PT cryostat. The temperature-dependent comparison between pristine and charged device states was performed in a Bluefors LD400 dilution refrigerator. Once a target temperature was reached the 2DEG in the Si QWs was first accumulated by applying to the gate electrode a DC gate voltage $V_G$ larger than the threshold voltage V$_{\text{thr}}$. The current was limited to approximately \(50~\mathrm{nA}\) in all measurements with a 1 M$\Omega$ preresistor. Hall measurements were then performed by sweeping the out-of-plane magnetic field \(B_\perp\), while recording the longitudinal voltage \(V_{xx}\), transverse voltage \(V_{xy}\), and source-drain current \(I_\mathrm{sd}\). For the temperature-dependent measurements in the reference-stack and deeper-QW sections, as well as for the \(1.5~\mathrm{K}\) benchmarking measurements, \(V_{xx}\) and \(V_{xy}\) were measured using a low-frequency (17 Hz) four-probe lock-in technique. In contrast, the temperature-dependent comparison between pristine and charged device states was performed using four-probe DC measurements. In this DC configuration, \(V_{xx}\) and \(V_{xy}\) were amplified using Basel Instruments SP1004 differential voltage amplifiers and the source-drain current was amplified using a Basel Instruments SP983c current-to-voltage converter. The amplified signals were read out using Keysight 34461A digital multimeters. In all measurements, the carrier density \(n\) was extracted from the Hall slope using
\[
\rho_{xy} = \frac{B_{\perp}}{en},
\]
where \(e\) is the elementary charge. The low-field mobility was calculated as
\[
\mu = \frac{1}{e n \rho_{xx}(0)},
\]
where \(n\) is the carrier density extracted from the low-field Hall slope and \(\rho_{xx}(0)\) is the longitudinal sheet resistivity evaluated near \(B_{\perp}=0\).  The longitudinal conductivity used for the percolation fit was calculated from the same zero-field longitudinal resistivity according to
\[
\sigma_{xx}(0) = \frac{1}{\rho_{xx}(0)}.
\]
The resulting \(\sigma_{xx}(0)\) was converted to units of \(e^2/h\) and used to extract the percolation density \(n_p\), as described in the main text.

\section*{Data availability}
The data supporting the findings of this study are publicly available on Zenodo at \href{doi.org/10.5281/zenodo.21930279}{(https://doi.org/10.5281/zenodo.21930279)}.

\section*{Acknowledgments}
Part of this work was supported by the German Federal Ministry of Research, Technology and Space within the project GeBaseQ. This work has been carried out within the Joint Lab “Spin-Based Quantum Computing” established between IHP - Leibniz Institute for High Performance Microelectronics, RWTH Aachen University and  Forschungszentrum Jülich.

\section*{Competing interests}
The authors declare no competing interests.

\section*{Author contributions}
\textbf{L. V.}:  Investigation, Formal analysis, Conceptualization, Visualization, Writing – original draft; 
\textbf{A. M.}:  Investigation, Formal analysis, Conceptualization, Visualization, Writing – original draft; 
\textbf{F. F.}: Writing – review \& editing, Investigation, Resources; 
\textbf{Y. Y.}: Writing – review \& editing, Resources;
\textbf{O. F.}: Writing – review \& editing, Investigation, Resources;
\textbf{S. M.}: Writing – review \& editing, Resources;
\textbf{M. H. Z.}: Writing – review \& editing;
\textbf{G. C.}: Writing – review \& editing;
\textbf{D. B.}: Writing – review \& editing;
\textbf{V. M.}: Writing – review \& editing, Funding acquisition, Project administration, Conceptualization, Resources;
\textbf{M. L.}: Writing – review \& editing, Investigation, Resources;
\textbf{F. R.:} Writing – review \& editing, Conceptualization, Investigation, Visualization, Formal analysis, Funding acquisition, Project administration

\bibliography{references}

\newpage

\setcounter{figure}{0}
\renewcommand{\thefigure}{S\arabic{figure}}

\clearpage
\onecolumngrid

\vspace*{1cm}

\begin{center}
    {\Large\bfseries
    Supplementary Online Material of\\
    Disorder Signatures Emerging at Millikelvin Temperatures\\
    in Si/SiGe Field-Effect Stacks
    }
\end{center}

\vspace{1cm}

\twocolumngrid

\section*{Reference-stack temperature dependence}
\subsection*{Full temperature sweep up to 10 K}

\begin{figure}[H]
    \centering
    \includegraphics[width=0.75\linewidth]{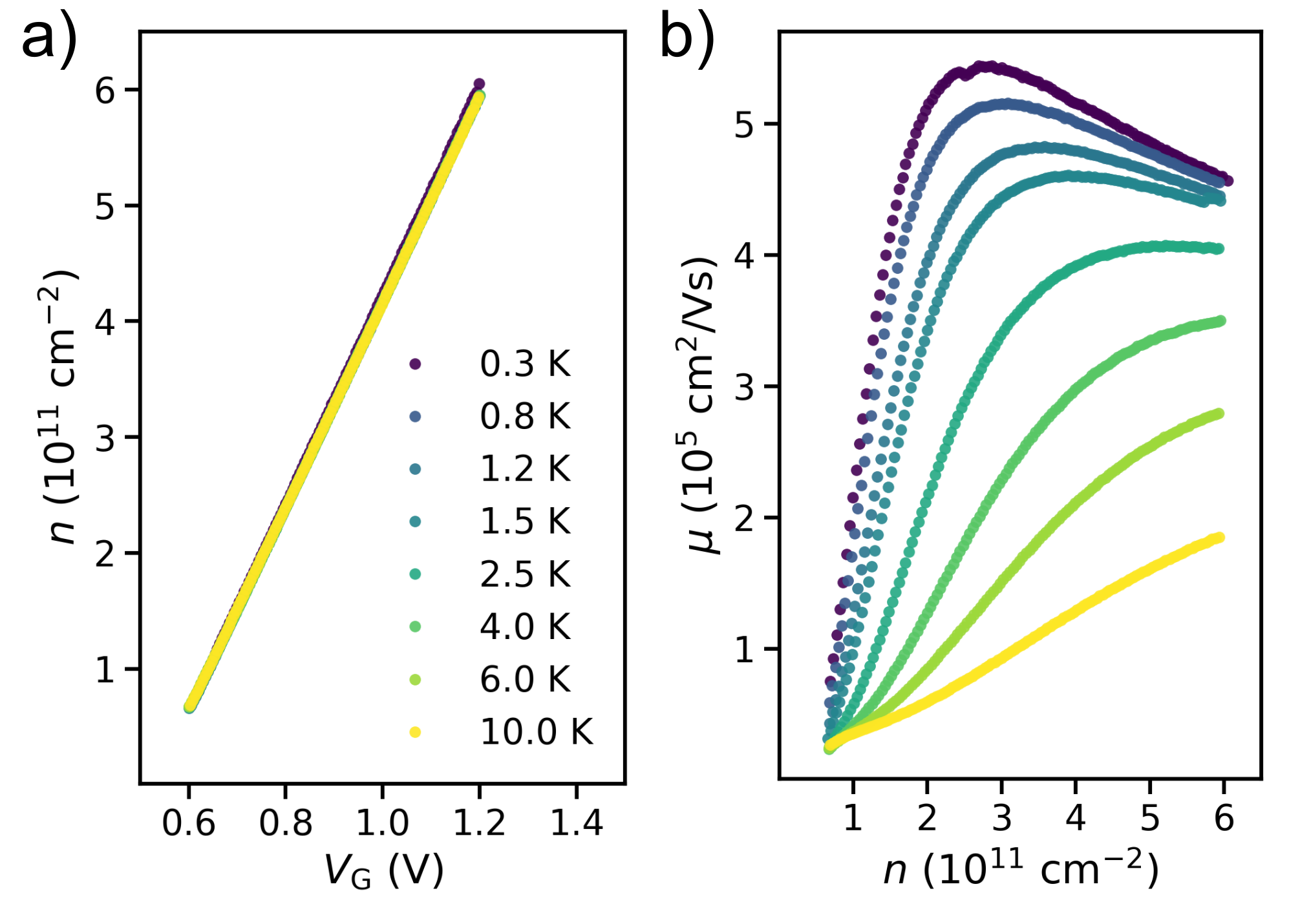}
        \caption{Temperature-dependent transport of the reference quantum-well stack.
        (a) Carrier density as a function of gate voltage \(n(V_\mathrm{G})\)measured between \(0.3~\mathrm{K}\) and \(10~\mathrm{K}\). The slope in the capacitive regime is largely temperature independent, indicating that the electrostatic gate coupling is not significantly modified over this temperature range.
        (b) Hall mobility as a function of carrier density \(\mu(n)\) for the same temperature range.}
    \label{SupFig_Reference_Stack_nvsVG_Mobvsn}
\end{figure}

Figure~\ref{SupFig_Reference_Stack_nvsVG_Mobvsn} shows the temperature dependence of the reference stack from $0.3~\mathrm{K}$ to $10~\mathrm{K}$. To minimize the influence of persistent charge configurations, the device was warmed up to room temperature for \(30~\mathrm{min}\) after each temperature point before being cooled down again for the subsequent measurement. This pristine device state then allows us to compare the temperature evolution of the density-dependent mobility. The $n(V_{\mathrm{G}})$ characteristics in panel~(a) remain nearly temperature independent, as indicated by the constant slope in the capacitively coupled regime. By contrast, the $\mu(n)$ curves in panel~(b) exhibit a pronounced temperature dependence, particularly near the onset of conduction. Cooling from $10~\mathrm{K}$ initially increases the mobility across the full density range. Below approximately $1.5~\mathrm{K}$, the high-density mobility changes only weakly, whereas the low-density mobility continues to increase down to $0.3~\mathrm{K}$.

\subsection*{Percolation threshold fitting}

\begin{figure}
    \centering
    \includegraphics[width=0.8\linewidth]{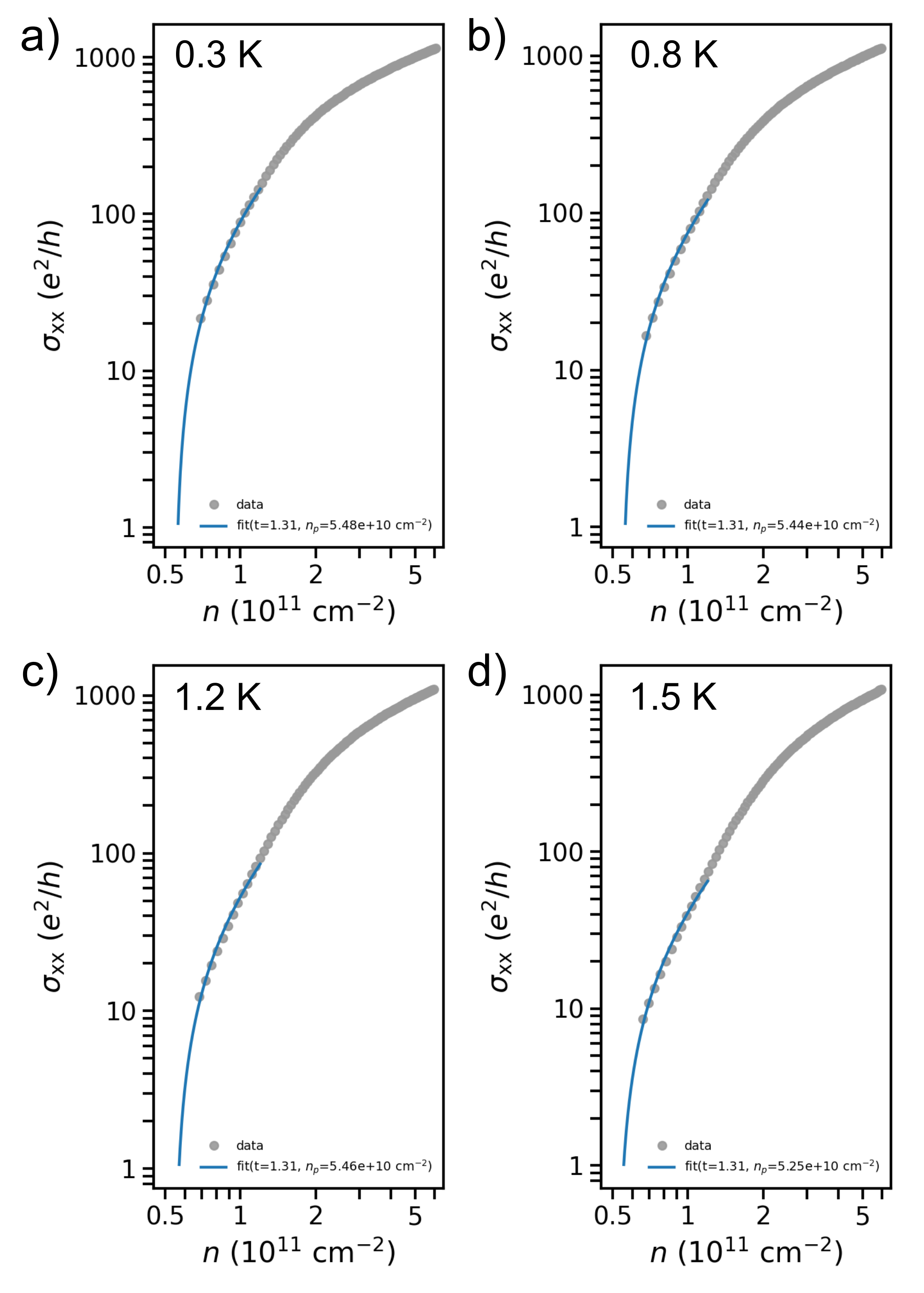}
        \caption{Extraction of the percolation density in the reference quantum-well stack.
        Longitudinal conductivity \(\sigma_{xx}\) as a function of carrier density \(n\) measured at
        (a) \(0.3~\mathrm{K}\), (b) \(0.8~\mathrm{K}\), (c) \(1.2~\mathrm{K}\), and (d) \(1.5~\mathrm{K}\).
        The solid lines show fits to the empirical percolation form
        \(\sigma_{xx} \propto (n-n_\mathrm{p})^\alpha\), used to extract the percolation density \(n_\mathrm{p}\).
        }
    \label{SupFig_Reference_Stack_Npfitting}
\end{figure}

Figure~\ref{SupFig_Reference_Stack_Npfitting} shows the longitudinal conductivity data and percolation fits for the reference stack at selected temperatures between $0.3~\mathrm{K}$ and $1.5~\mathrm{K}$. The data are fitted using $\sigma_{xx} \propto (n-n_{\mathrm{p}})^\alpha$, where $n_{\mathrm{p}}$ denotes the critical density for the onset of extended conduction. Within the fitting uncertainty, $n_{\mathrm{p}}$ remains nearly temperature independent, indicating that the low-temperature mobility changes do not arise from a shift in the threshold itself.

\section*{Interface-trap filling and charge-history effects}
\subsection*{Charging behavior}

\begin{figure}
    \centering
    \includegraphics[width=0.8\linewidth]{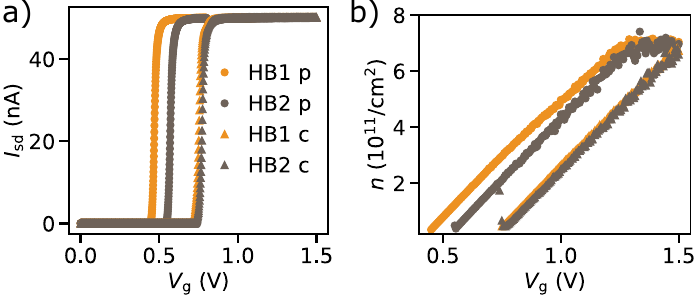}
        \caption{Charging behavior of the HB1 and HB2 devices measured at $0.2~\mathrm{K}$. (a) Turn-on characteristics in the pristine (p) and charged (c) states. (b) Carrier density as a function of gate voltage beyond the purely capacitive regime.}
    \label{fig:iv_charging}
\end{figure}

Figure~\ref{fig:iv_charging} compares the charging behavior of the two HB-FET devices, HB1 and HB2, measured at $0.02~\mathrm{K}$. The pristine and charged states are denoted by ''p'' and ''c'', respectively. As shown in panel~(a), the pristine devices exhibit different turn-on voltages, consistent with device-to-device variations in their initial electrostatic environment. After charging, however, their turn-on characteristics become closely aligned. The corresponding carrier-density characteristics in panel~(b) show similar slopes in the pristine state, indicating comparable gate capacitances, while their horizontal offset reflects the difference in threshold voltage. Following charging, the $n(V_\mathrm{G})$ curves nearly overlap, suggesting that the charging procedure establishes similar electrostatic boundary conditions and effective charge distributions at the dielectric interface in both devices.

\subsection*{Temperature dependence of post-charging device characteristics without thermal reset}

\begin{figure}
    \centering
    \includegraphics[width=0.8\linewidth]{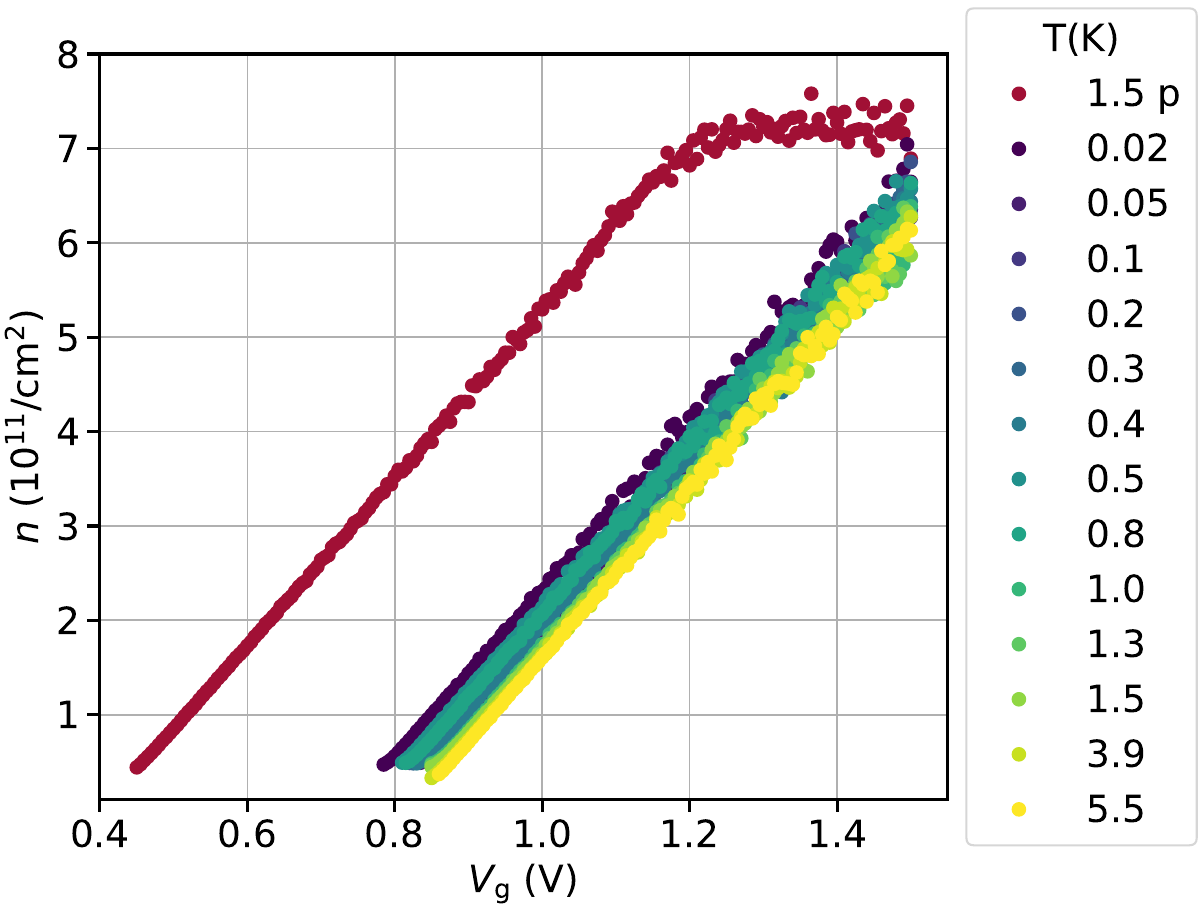}
        \caption{Carrier density $n$ as a function of gate voltage $V_{\mathrm{G}}$ during the post-charging temperature-dependent characterization. The initial measurement in the pristine state is shown in red. During the first sweep at $0.02~\mathrm{K}$, the gate voltage was increased up to $1.5~\mathrm{V}$, resulting in a shift of the $n(V_{\mathrm{G}})$ characteristic.}
        \label{fig:charged_device_n_vs_vg}
\end{figure}

Figure~\ref{fig:charged_device_n_vs_vg} shows the carrier density as a function of gate voltage during the temperature-dependent characterization. The initial measurement, shown in red, represents the pristine device state. During the first sweep at $0.02~\mathrm{K}$, the gate voltage was increased up to $1.5~\mathrm{V}$, producing a horizontal shift of the $n(V_{\mathrm{G}})$ characteristic that is consistent with charging of the device. After this initial sweep, the $n(V_{\mathrm{G}})$ curves measured at the subsequent temperatures remain essentially unchanged. This indicates that the device reaches a comparatively stable post-charging state and that no pronounced additional threshold-voltage shift occurs during the remaining temperature sequence. Although shifts in the turn-on voltage were observed after the gate voltage was first increased to $1.5~\mathrm{V}$, this value was maintained as the upper limit in all subsequent measurements. The non-monotonic sequence of measurement temperatures is summarized in Table~1.

\begin{figure}
    \centering
    \includegraphics[width=0.85\linewidth]{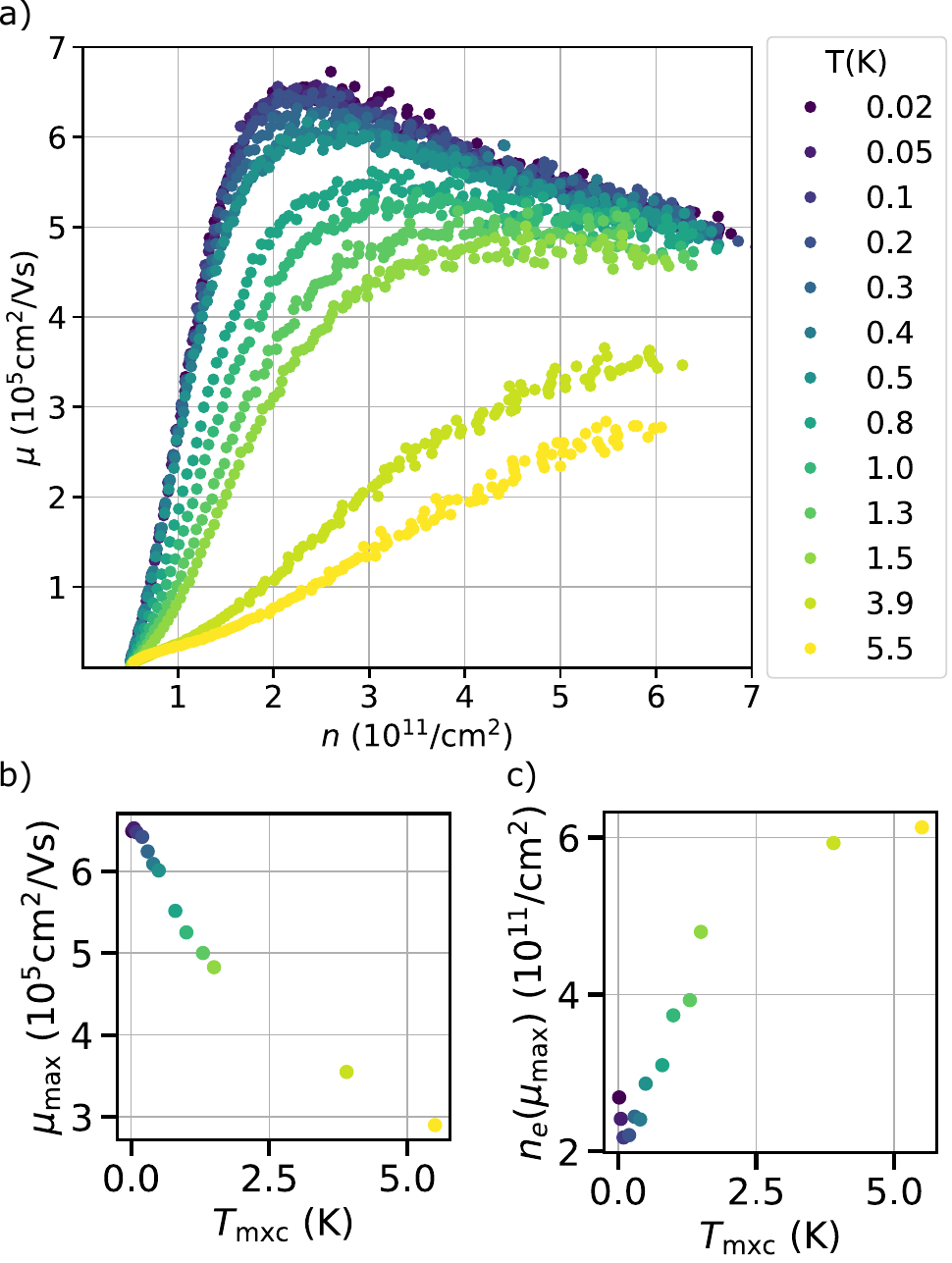}
        \caption{Temperature dependence of the mobility characteristics in the post-charging state without an intermediate thermal reset. (a) Hall mobility $\mu$ as a function of carrier density $n$, measured between $5.5~\mathrm{K}$ and $0.02~\mathrm{K}$. For each measurement, the gate voltage was swept up to $1.5~\mathrm{V}$, while the device remained in the post-charging state throughout the non-monotonic temperature sequence. (b) Maximum mobility $\mu_{\mathrm{max}}$ extracted from the smoothed $\mu(n)$ curves. (c) Carrier density $n(\mu_{\mathrm{max}})$ at which the mobility maximum occurs.}
        \label{fig:charged_device_temperature_dependence}
\end{figure}

Figure~\ref{fig:charged_device_temperature_dependence}(a) shows the Hall mobility as a function of carrier density for measurement temperatures between $0.02~\mathrm{K}$ and $5.5~\mathrm{K}$ and according to sweep order of Table~1. The qualitative temperature dependence resembles that observed for the pristine device state discussed in the first section of the main text. Above approximately $1.5~\mathrm{K}$, cooling increases the mobility across most of the investigated carrier-density range. Below this temperature, the changes become concentrated primarily in the low-density regime, whereas the high-density mobility varies only weakly. Figure~\ref{fig:charged_device_temperature_dependence}(b) shows that the maximum mobility increases continuously with decreasing temperature. At the same time, Figure~\ref{fig:charged_device_temperature_dependence}(c) reveals a systematic shift of the corresponding carrier density toward lower values. Cooling therefore not only increases the maximum mobility but also shifts the mobility maximum progressively closer to the low density regime.

\begin{table}[t]
\caption{\label{tab:measurement_sequence}%
Measurement order and corresponding mixing-chamber temperatures for the classical Hall characterization.}
\centering
    \begin{ruledtabular}
        \begin{tabular}{c c c c c c c c}
            Run
            & 1 & 2 & 3 & 4 & 5 & 6 & 7 \\
            $T_{\mathrm{mxc}}$ ($\mathrm{K}$)
            & 0.02 & 0.5 & 0.8 & 0.2 & 0.1 & 0.3 & 0.4 \\
            \hline
            Run
            & 8 & 9 & 10 & 11 & 12 & 13 & {} \\
            $T_{\mathrm{mxc}}$ ($\mathrm{K}$)
            & 0.05 & 1.0 & 1.5 & 1.3 & 3.9 & 5.5 & {} \\
        \end{tabular}
    \end{ruledtabular}
\end{table}

\subsection*{Consistency of the pristine state across cooldowns}

\begin{figure}
    \centering
    \includegraphics[width=0.8\linewidth]{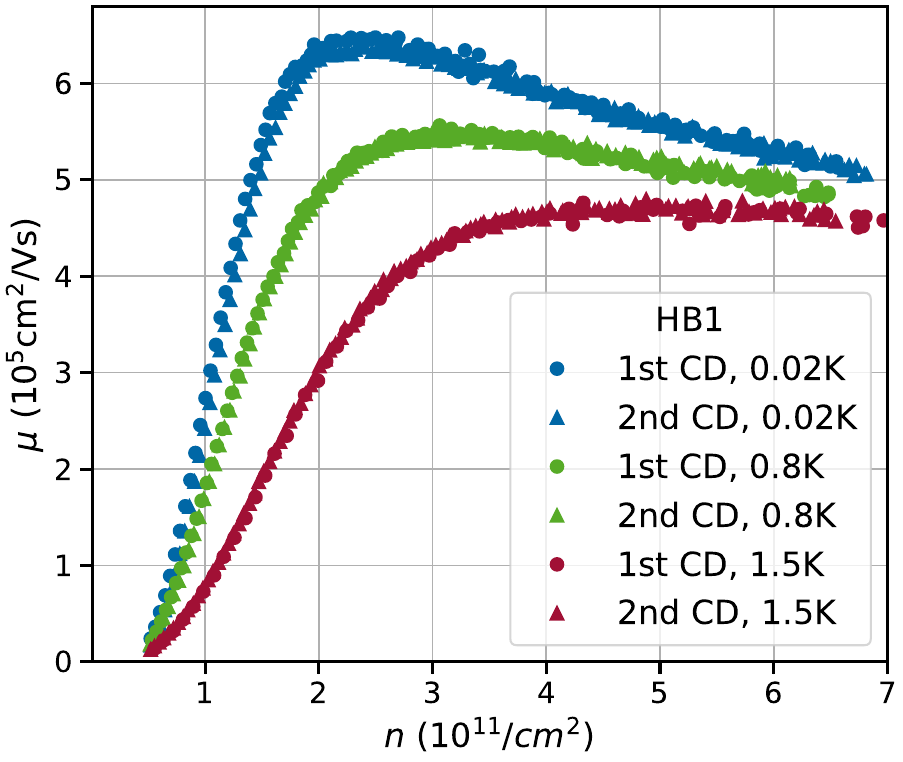}
        \caption{Reproducibility of the pristine-state mobility characteristics across independent thermal cycles. Hall mobility $\mu$ as a function of carrier density $n$ measured during two separate cooldowns at $0.02~\mathrm{K}$, $0.8~\mathrm{K}$, and $1.5~\mathrm{K}$. Before each cooldown, the device was returned to room temperature to thermally reset the charge state. The close agreement between the two measurements at each temperature indicates that the pristine transport state is reproducibly established after thermal cycling.}
    \label{fig:multiple_cooldowns_pristine}
\end{figure}

To assess whether the initial transport state is reproducibly established after thermal cycling, the device was measured during two independent cooldowns at each of the three temperatures considered, $0.02~\mathrm{K}$, $0.8~\mathrm{K}$, and $1.5~\mathrm{K}$. Before every cooldown, the device was returned to room temperature, thereby thermally resetting the charge state, and each measurement was subsequently performed on the freshly cooled device in its pristine state. As shown in Figure~\ref{fig:multiple_cooldowns_pristine}, the corresponding $\mu(n)$ characteristics obtained during the two cooldowns at each temperature agree closely, with no significant shift or systematic modification. This reproducibility indicates that the initial transport state is largely preserved across independent thermal cycles and that the observed temperature dependence is not dominated by stochastic cooldown-to-cooldown variations in the electrostatic environment.

\subsection*{Charge-history-dependent mobility of HB2}

\begin{figure}
    \centering
    \includegraphics[width=0.8\linewidth]{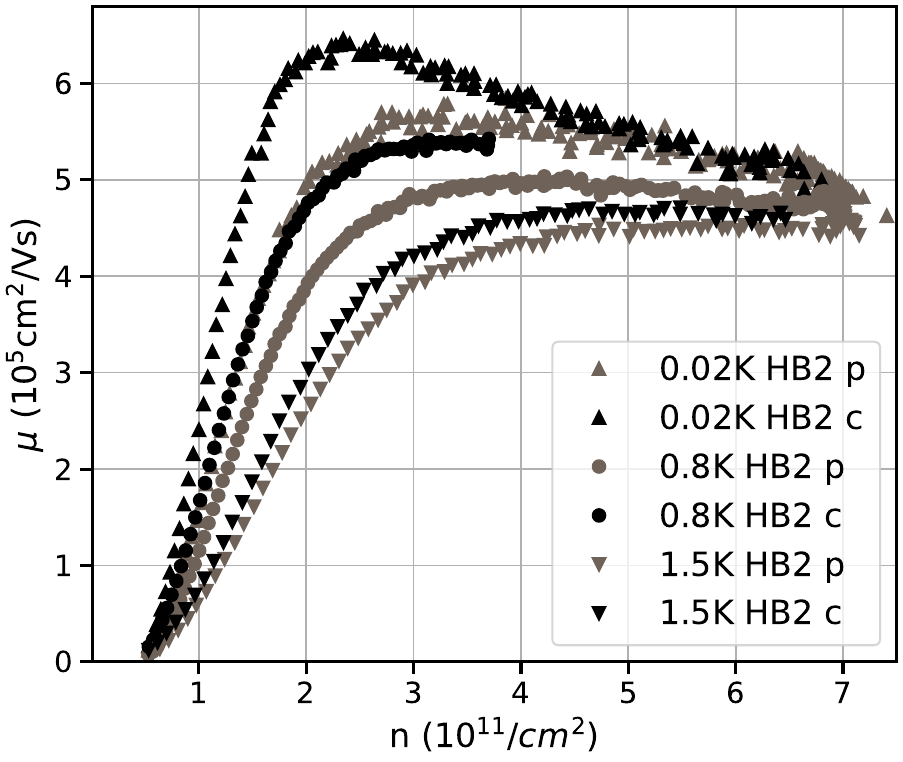}
        \caption{Charge-history-dependent mobility of HB2. Hall mobility $\mu$ as a function of carrier density $n$ in the pristine and charged states at the three investigated mixing-chamber temperatures. The charging-induced mobility enhancement is strongest at $0.02~\mathrm{K}$ and decreases with increasing temperature, reproducing the qualitative behavior observed for HB1.}
    \label{fig:hb2_charging}
\end{figure}

To examine whether the charge-history dependence observed for HB1 is also present in a second device, HB2 was characterized in both the pristine and charged state at the three investigated mixing-chamber temperatures. At each temperature, the device was first measured after a thermal reset to room temperature and was subsequently remeasured following the high gate bias charging sweep.

Figure~\ref{fig:hb2_charging} shows that HB2 exhibits the same qualitative behavior as HB1. Charging produces the largest mobility enhancement at $0.02~\mathrm{K}$, particularly in the low- to intermediate-density regime, while the effect becomes progressively weaker with increasing temperature. At $1.5~\mathrm{K}$, only a minor difference between the pristine and charged states remains. The agreement between the two devices demonstrates that the low-temperature mobility enhancement following a high-bias sweep is not specific to HB1, but represents a reproducible charge-history-dependent transport response of the investigated devices.

\section*{Temperature-dependent transport in deep quantum wells}
\subsection*{Carrier density dependent mobility including intermediate temperature sweeps}

\begin{figure}
    \centering
    \includegraphics[width=0.95\linewidth]{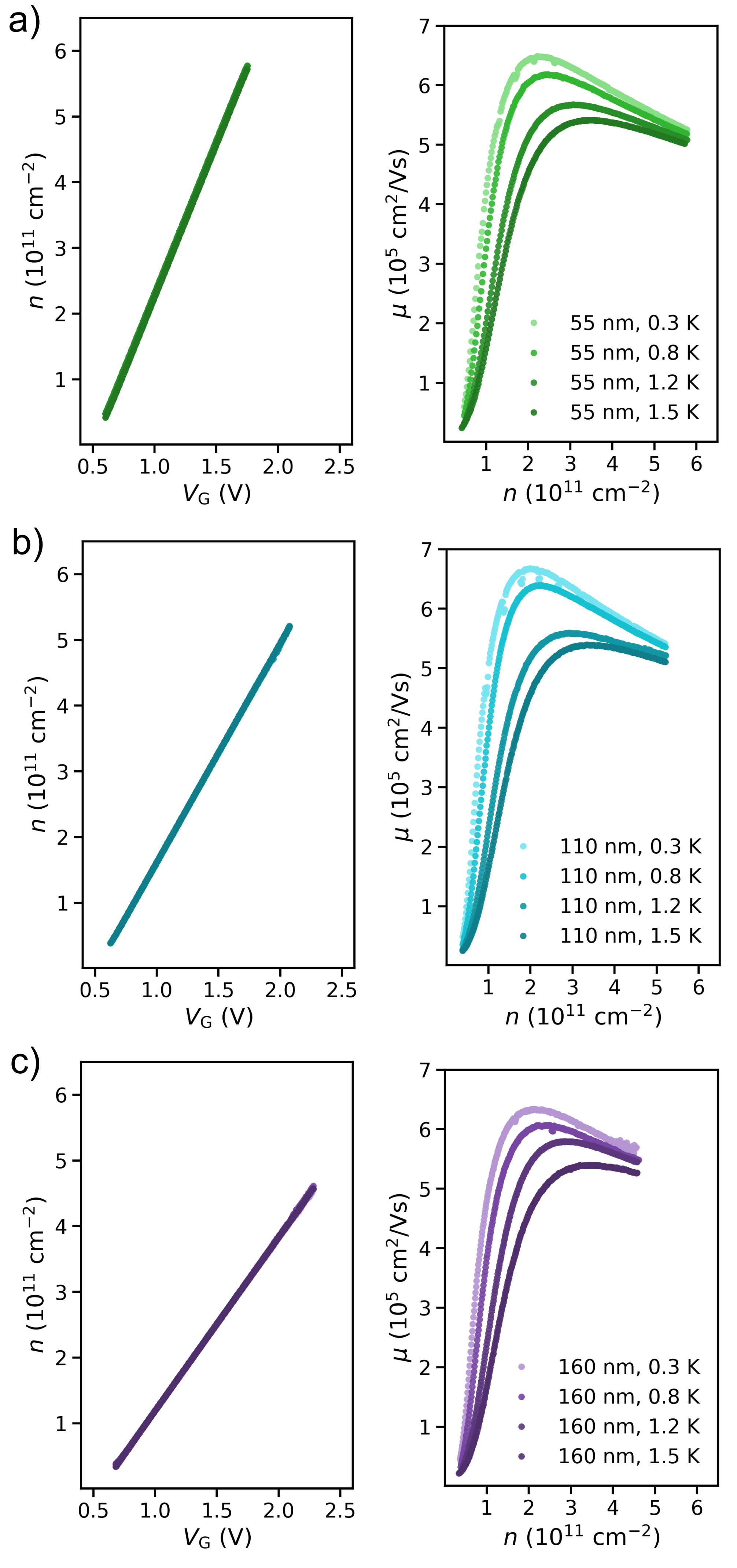}
        \caption{Temperature-dependent transport in deeper quantum-well stacks.
        Carrier density as a function of gate voltage \(n(V_\mathrm{G})\) and Hall mobility as a function of carrier density \(\mu(n)\) are shown for quantum-well depths of (a) \(55~\mathrm{nm}\), (b) \(110~\mathrm{nm}\), and (c) \(160~\mathrm{nm}\).
        For each quantum-well depth, the left panel shows \(n(V_\mathrm{G})\) and the right panel shows \(\mu(n)\), measured at \(1.5~\mathrm{K}\), \(1.2~\mathrm{K}\), \(0.8~\mathrm{K}\), and \(0.3~\mathrm{K}\), indicated by a brightness scale.}
    \label{SupFig_DeepQW_nvsVG_Mobvsn}
\end{figure}

Figure~\ref{SupFig_DeepQW_nvsVG_Mobvsn} summarizes the full temperature-dependent transport data for the deeper quantum-well stacks with QW depths of a) \(55~\mathrm{nm}\), b) \(110~\mathrm{nm}\), and c) \(160~\mathrm{nm}\). For all three stacks, \(n(V_\mathrm{G})\) (left panel) shows no significant change in slope or shifting along \(V_\mathrm{G}\) between \(1.5~\mathrm{K}\) and \(0.3~\mathrm{K}\). This indicates that the electrostatic gate coupling in the considered density range is not significantly modified upon cooling. 

The mobility curves shown in the right panel show a systematic enhancement upon cooling for each QW depth. As in the reference stack, the temperature dependence is most pronounced at low  and intermediate carrier density.At higher density, the mobility changes more weakly between \(1.5~\mathrm{K}\) and \(0.3~\mathrm{K}\), indicating that this regime is closer to its low-temperature limit already at \(1.5~\mathrm{K}\). The full temperature sweeps therefore support the main-text conclusion that the deep-QW stacks exhibit stable electrostatics but retain a density-dependent temperature sensitivity in their mobility. 

\subsection*{Fractional mobility including intermediate temperature sweeps}

\begin{figure*}[t]
    \centering
    \includegraphics[width=0.8\textwidth]{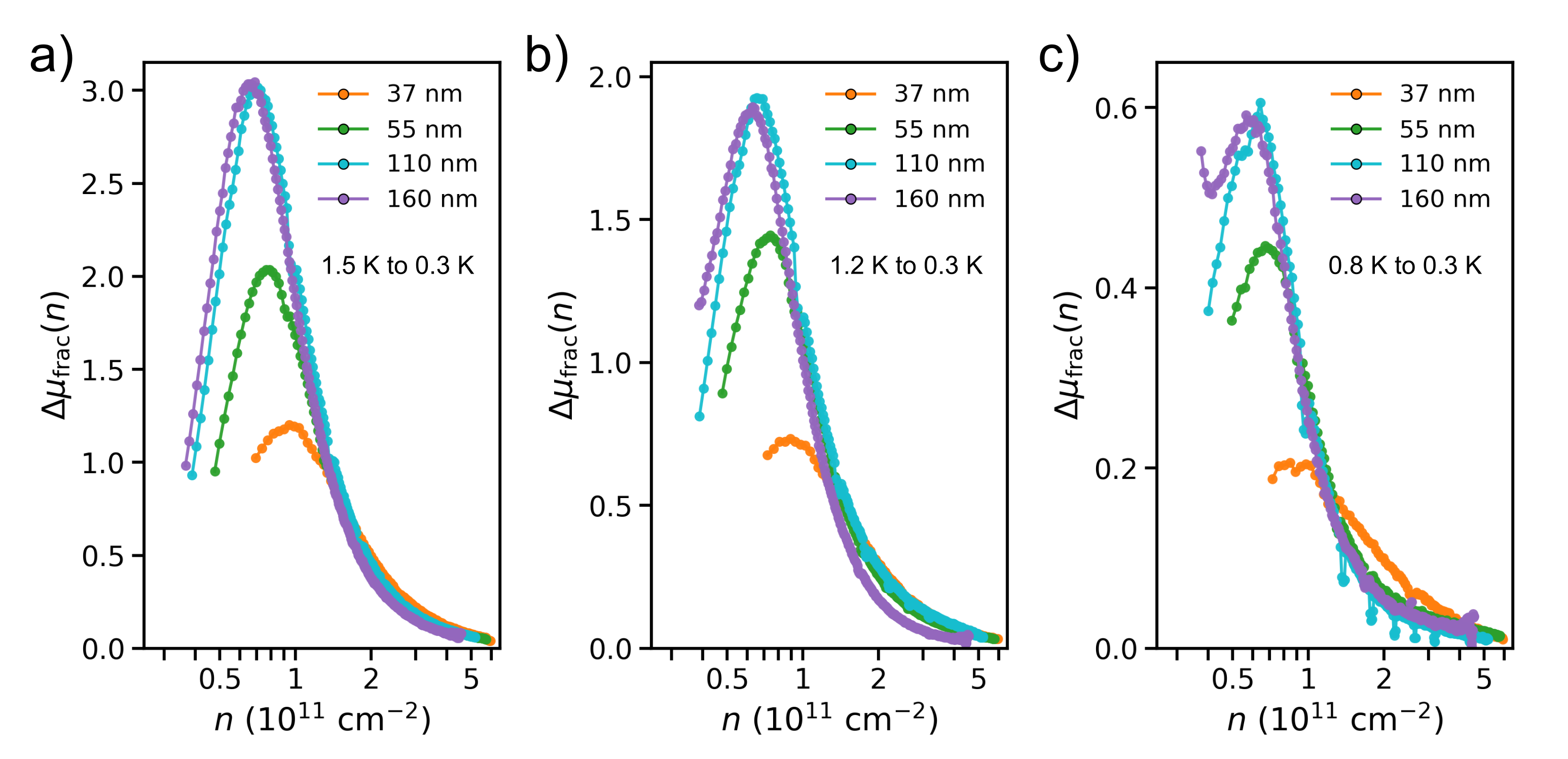}
        \caption{Comparison of the fractional mobility gain for different quantum-well depths and temperature intervals.
        The fractional mobility gain is shown for quantum-well depths of \(37~\mathrm{nm}\), \(55~\mathrm{nm}\), \(110~\mathrm{nm}\), and \(160~\mathrm{nm}\), calculated between (a) \(1.5~\mathrm{K}\) and \(0.3~\mathrm{K}\), (b) \(1.2~\mathrm{K}\) and \(0.3~\mathrm{K}\), and (c) \(0.8~\mathrm{K}\) and \(0.3~\mathrm{K}\).
        }
    \label{SupFig_DeepQW_160_nvsVG_Mobvsn}
\end{figure*}

Figure~\ref{SupFig_DeepQW_160_nvsVG_Mobvsn} compares the fractional mobility gain for the four quantum-well depths using three different temperature intervals, namely a) \(1.5~\mathrm{K}\) to \(0.3~\mathrm{K}\), b) \(1.2~\mathrm{K}\) to \(0.3~\mathrm{K}\)  and c) \(0.8~\mathrm{K}\) to \(0.3~\mathrm{K}\). In each case, the same qualitative behavior is observed: the fractional gain rises above the onset of conduction, reaches a maximum at low to intermediate carrier density, and then decreases toward higher density. Most importantly, the relative ordering with quantum-well depth remains unchanged. The \(110~\mathrm{nm}\) and \(160~\mathrm{nm}\) stacks consistently exhibit the largest low-density fractional gain, followed by the \(55~\mathrm{nm}\) stack and the \(37~\mathrm{nm}\) reference stack. At the same time, the overall magnitude of the fractional mobility gain decreases systematically as the higher comparison temperature is moved from \(1.5~\mathrm{K}\) to \(1.2~\mathrm{K}\) and then to \(0.8~\mathrm{K}\). The persistence of the depth-dependent ordering despite this reduced amplitude supports the interpretation of the main text that the fractional mobility gain captures an intrinsic difference in the temperature-sensitive low-density transport regime. Overall, these results demonstrates that the trend identified in the main text is not tied to that specific temperature pair, but instead reflects a feature of the low-temperature transport evolution.

\section*{Methods: Magnetotransport Characterization}

The measurements at FZJ were performed in a Bluefors LD400 dilution refrigerator equipped with custom DC filtering and a $9$--$1$--$1~\mathrm{T}$ vector magnet. The DC filter stages comprised RC filters with $R_1 = 510~\Omega$ (ERA6AED511V), $R_2 = 1.2~\mathrm{k}\Omega$ (ERA6VEB1201V), $C_1 = 0.068~\mu\mathrm{F}$ (C0603X683K4RACAUTO), and $C_2 = 0.39~\mu\mathrm{F}$ (C0603C394K4RACAUTO), as well as LC filters (TDK MEM1608P25R0T001, TDK MEM1608P75R0T001, Mini-Circuits LFCN-1700+, and Mini-Circuits LFCN-5000+). The samples were mounted on chip carriers such that the main magnetic-field axis was aligned with the out-of-plane direction of the devices.

As for the measurements described in the previous section, a four-terminal configuration was used. Here, however, the measurements were performed using DC excitation rather than lock-in detection. The longitudinal and transverse voltages, $V_{xx}$ and $V_{xy}$, were amplified using Basel Instruments SP1004 differential voltage amplifiers with a low-pass cutoff frequency of $f_{\mathrm{cutoff}} = 300~\mathrm{Hz}$. The gains were set to $g_{xx} = 10^{2}$ and $g_{xy} = 10^{3}$, respectively. The source--drain current, $I_{\mathrm{sd}}$, was amplified using a Basel Instruments SP983c current-to-voltage converter with a low-pass cutoff frequency of $f_{\mathrm{cutoff}} = 300~\mathrm{Hz}$ and a gain of $g_{\mathrm{sd}} = 10^{7}~\mathrm{V/A}$. The resulting output voltages were recorded using Keysight 34461A digital multimeters. The gate voltage was supplied by a QDevil QDAC-II. The same instrument was used to apply a source--drain bias voltage of $50~\mathrm{mV}$ through a $1~\mathrm{M}\Omega$ series resistor, thereby limiting $I_{\mathrm{sd}}$ to $50~\mathrm{nA}$.

\begin{figure}
    \centering
    \includegraphics[width=0.8\linewidth]{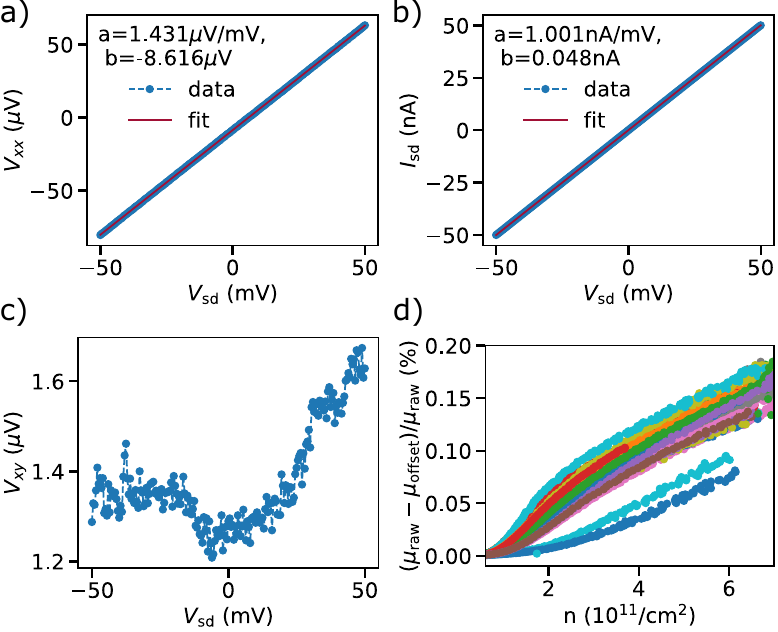}
        \caption{Exemplary extraction of the DC amplifier offsets. Source--drain-bias dependence of (a) the source--drain current, (b) the longitudinal voltage drop, and (c) the Hall voltage. (d) Fractional difference between the offset-corrected and uncorrected mobilities for all Hall measurements performed at FZJ and included in this study.}
    \label{fig:offset_correction}
\end{figure}

The carrier density and mobility were determined using the same procedure as for the measurements performed at IHP. To compensate for slowly drifting DC offsets in the differential amplifiers, the source--drain bias was swept between $-50~\mathrm{mV}$ and $+50~\mathrm{mV}$ at a conductive operating point within the linear $n(V_{\mathrm{g}})$ regime. An example of this procedure is shown in Fig.~\ref{fig:offset_correction}(a)--(c). Linear fits to $I_{\mathrm{sd}}(V_{\mathrm{sd}})$ and $V_{xx}(V_{\mathrm{sd}})$ were used to determine the respective offsets from the $y$-intercepts. The $V_{xy}$ offset was determined as the mean voltage over the full source--drain-bias range.

For all datasets included in this study, the fractional difference between the mobilities obtained with and without offset correction was calculated, as shown in Fig.~\ref{fig:offset_correction}(d). The magnitude of the correction follows the overall density dependence of $\mu(n)$ because the longitudinal voltage drop decreases with increasing conductivity. The correction remains below $0.2\,\%$ for all datasets and consistently results in a slightly lower extracted mobility. 

\end{document}